\documentclass[useAMS, usenatbib]{mn2e} 
\usepackage{aas_macros}
\usepackage{graphicx}
\usepackage{subfig}
\usepackage{amsmath}
\usepackage{pdflscape}
\usepackage{fixltx2e}
\usepackage{longtable}
\usepackage{color}
\long\def\symbolfootnote[#1]#2{\begingroup%
\def\thefootnote{\fnsymbol{footnote}}\footnote[#1]{#2}\endgroup}

\title{WISE Circumstellar Disks in the Young Sco-Cen Association}
\author[A.C. Rizzuto et al.]{A.C. Rizzuto$^{1}$, M.J. Ireland$^{1,2}$, D.B. Zucker$^{1,2}$\\
$^{1}$Department of Physics and Astronomy, Macquarie University, Sydney NSW, 2109, Australia\\
$^{2}$Australian Astronomical Observatory, Epping NSW 2121, Australia}

\newcommand{\mic}{$\mu$m~}
\newcommand{\jk}{(J-K$_s$)~}
\newcommand{\ws}{W$_1$-W$_2$}
\newcommand{\whm}{W$_{1}$-W$_3$}
\newcommand{\whl}{W$_{1}$-W$_4$}

\begin{document}

\pagerange{\pageref{firstpage}--\pageref{lastpage}} \pubyear{2011}

\maketitle
\begin{abstract}
We present an analysis of the WISE photometric data for 829 stars in the Sco-Cen OB2 association, using the latest high-mass membership probabilities. We detect infrared excesses associated with 135 BAF-type stars, 99 of which are secure Sco-Cen members. There is a clear increase in excess fraction with membership probability, which can be fitted linearly. We infer that 41$\pm$5\% of Sco-Cen OB2 BAF stars to have excesses, while the field star excess fraction is consistent with zero. This is the first time that the probability of non-membership has been used in the calculation of excess fractions for young stars. We do not observe any significant change in excess fraction between the three subgroups. Within our sample, we have observed that B-type association members have a significantly smaller excess fraction than A and F-type association members.
\end{abstract}


\section{Introduction}
Many young ($\sim$1-10\,Myr) stars of all types, ranging from tens of solar masses down to the smallest brown dwarves and in all environments, are surrounded by circumstellar disks \citep{strom89,lada2000,carpenter06}. A variety of observations have provided us with an overall timeline of disk evolution. The inner portion of the disk ($\leq 1$\,AU) dissipates by the age of 10\,Myr in all but a small fraction of stars \citep{mamajek04,silverstone06}. From the age of 10\,Myr onwards, there is an observed decline in the 24$\mu$ excess relative to the photosphere for stars of B to K type \citep{gaspar09,rieke05}. It has been postulated that planetesimal stirring though stellar ages of 5 to 20\,Myr could produce an increase in the strength of the $\sim$24\mic excess with age \citep{kenyon08}, however the current data do not show a statistically significant increase \citep{carpenter09}. 

The Sco-Cen association and its three subgroups-Upper Scorpius (US), Upper Centaurus Lupus (UCL) and Lower Centaurus Crux (LCC)-provide three important constant-age samples within which to study circumstellar disks around young stars. The subgroups have ages of $\sim$5, $\sim$16, and $\sim$17\,Myr respectively, and are located less than 150\,pc from the Sun \citep{zeeuw99}. 
Previous studies \citep{carpenter06,carpenter09} have investigated IRAC, IRS and Spitzer photometric data ranging from 4.5 to 70\,\mic in the US subgroup. They identified 54 stars with 24\mic excesses in their sample of 205 targets and found that disks around BAF-type stars appear to be comprised of dusty debris, while disks associated with K and M-type stars are likely optically thick primordial disks which are remnants of the star formation process. 
The older Sco-Cen subgroups, UCL and LCC have received considerably less attention; only a handful of studies have observed small numbers of UCL and LCC stars \citep{su06,carpenter09}. The recent study by \citet{chen11} reports the detection of 41 new disks around F and G-type Sco-Cen stars.

New Bayesian membership probabilities for the high mass members (B to F-type) of the Sco-Cen association are now available \citep{myfirstpaper}, and preliminary photometric data from the WISE mission have been released \citep{wise10}. In this paper we present an analysis of the WISE photometry for Sco-Cen members to search for circumstellar disks in three constant-age samples.

\section{Data Sample}
In this study we take our sample from the Hipparcos Sco-Cen membership study of \citet{myfirstpaper}. All stars with membership probabilities of 5\% and greater were cross-referenced with the WISE preliminary data release. This resulted in a sample of 829 stars with spectral types ranging from early B to late F (B-V$\leq$0.6),  brighter than 9$^{th}$ visual magnitude and within the area of sky bounded by ($285 \leq l \leq 360$) and ($-10 \leq b \leq 60$). Taking into consideration membership probabilities, this equates to $\sim$400 true members. Membership as described in \citet{myfirstpaper} is based purely on kinematic and positional properties of the targets, and hence the selection is believed to be unbiased with regard to the presence of circumstellar disks. We have chosen to include the low membership probability stars in this study in order to use a potential relationship between excess fraction and probability of membership to infer a more accurate excess fraction for secure  members.

\section{WISE  Excesses}

The WISE mission data provides photometry in four bands, W$_1$, W$_2$, W$_3$ and W$_4$, with central wavelengths of 3, 4.5, 12 and 22\,\mic respectively \citep{wise10}. In this study we present an analysis of three WISE colours: \ws, \whm, and \whl. Inspection of the WISE photometry for band W$_2$ shows a clear bias at the bluest end of our membership sample towards poorly fitted point-spread-functions, resulting in untrustworthy W$_2$ photometry. For this reason the long wavelength colours were constructed with W$_1$, which shows a uniform distribution across spectral type of poor photometry fits. Analysis of these three colours will provide three different classes of detected excesses: (1) excesses in all colours, (2) excesses in only the short wavelength filters, and (3) excess in only the longer or longest wavelengths. It is expected that a disk-produced excess detected at bluer colours will also be detected at longer colours, hence, excesses in only the blue colours will indicate contaminated photometry.

Stellar photospheric emission is expected to vary linear with 2MASS \jk colour. To determine the photospheric colours, we have applied an iterative fitting procedure. Objects which were clear outliers in the particular WISE colour were first removed and then the software package {\sc mpfit} was used to fit a line to the sample. Objects separated from the fitted line by more than twice the dispersion of the residuals were then removed as outliers. The process was then repeated until no further objects were removed. This fitting procedure was carried out using only the stars with greater than 60\% membership probability in order to ensure that the fitted photosphere line was that of the young Sco-Cen stars. During this fitting procedure, objects with a WISE photometric fit reduced $\chi^2$ greater than 4 in the relevant bands were excluded, as they are likely to be extended sources with poor photometry. 
\begin{figure}
\subfloat[\ws]{\label{w12}\includegraphics[width=0.5\textwidth]{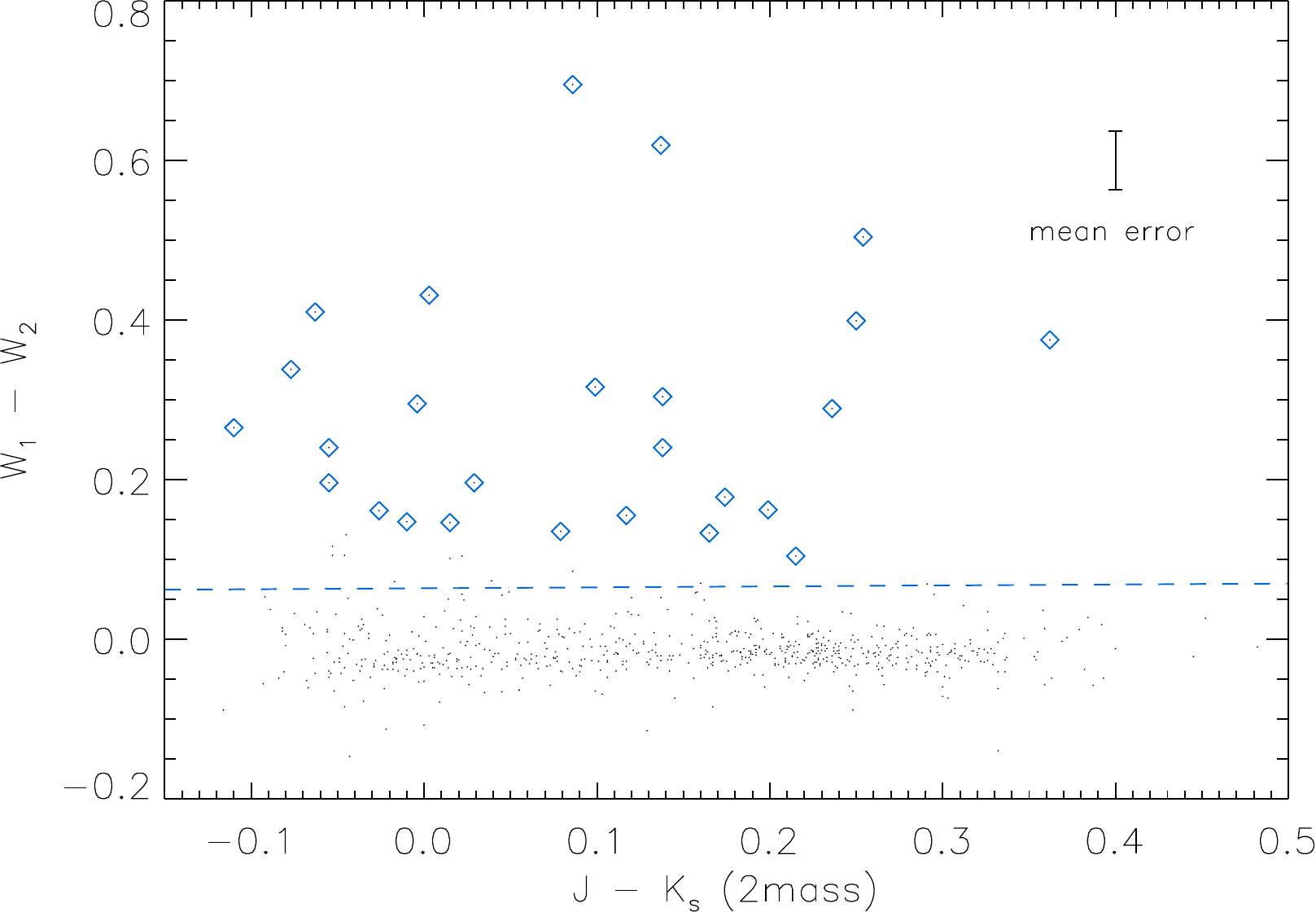}}\\
\subfloat[\whm]{\label{wh3}\includegraphics[width=0.5\textwidth]{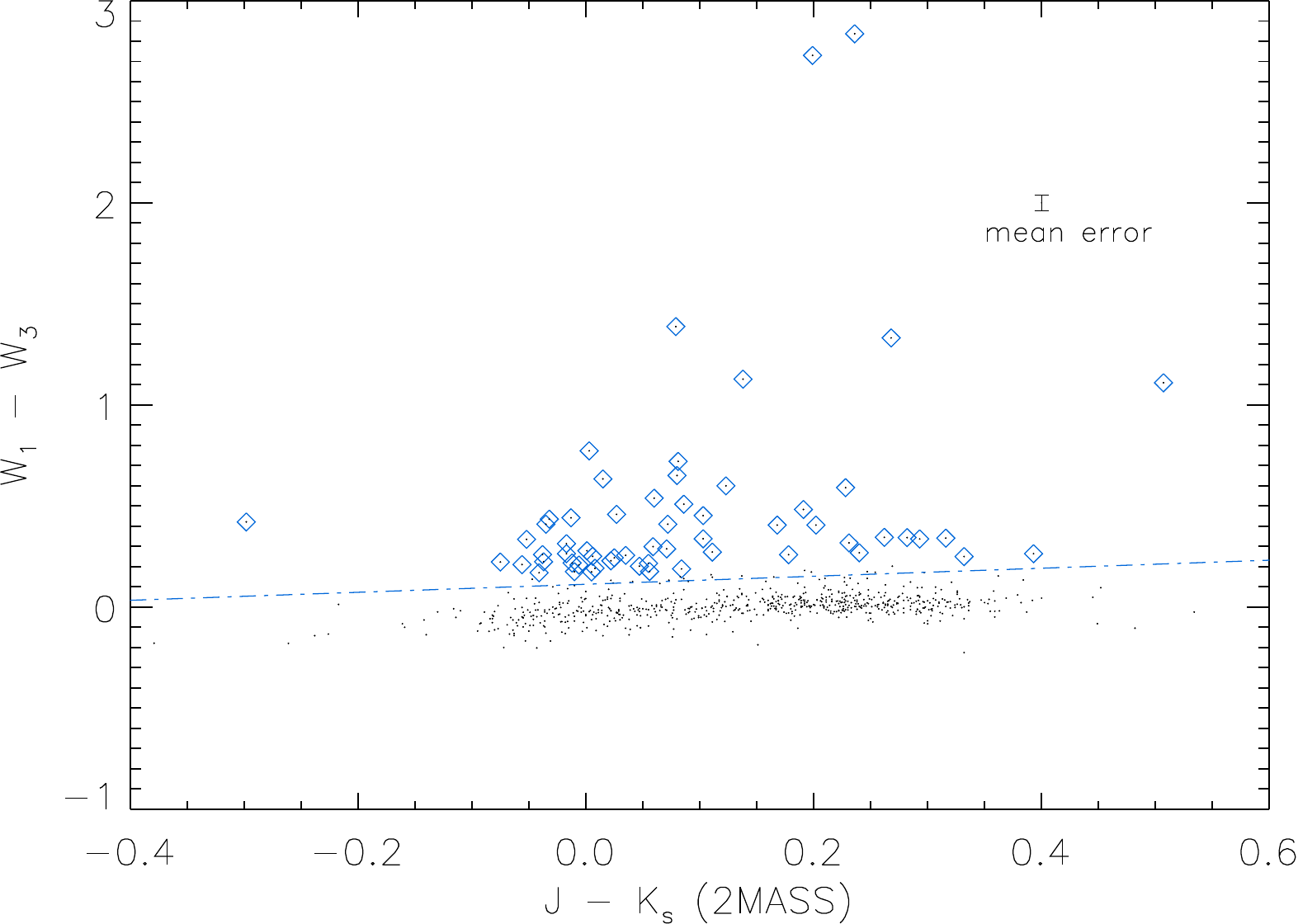}}\\
\subfloat[\whl]{\label{wh4}\includegraphics[width=0.5\textwidth]{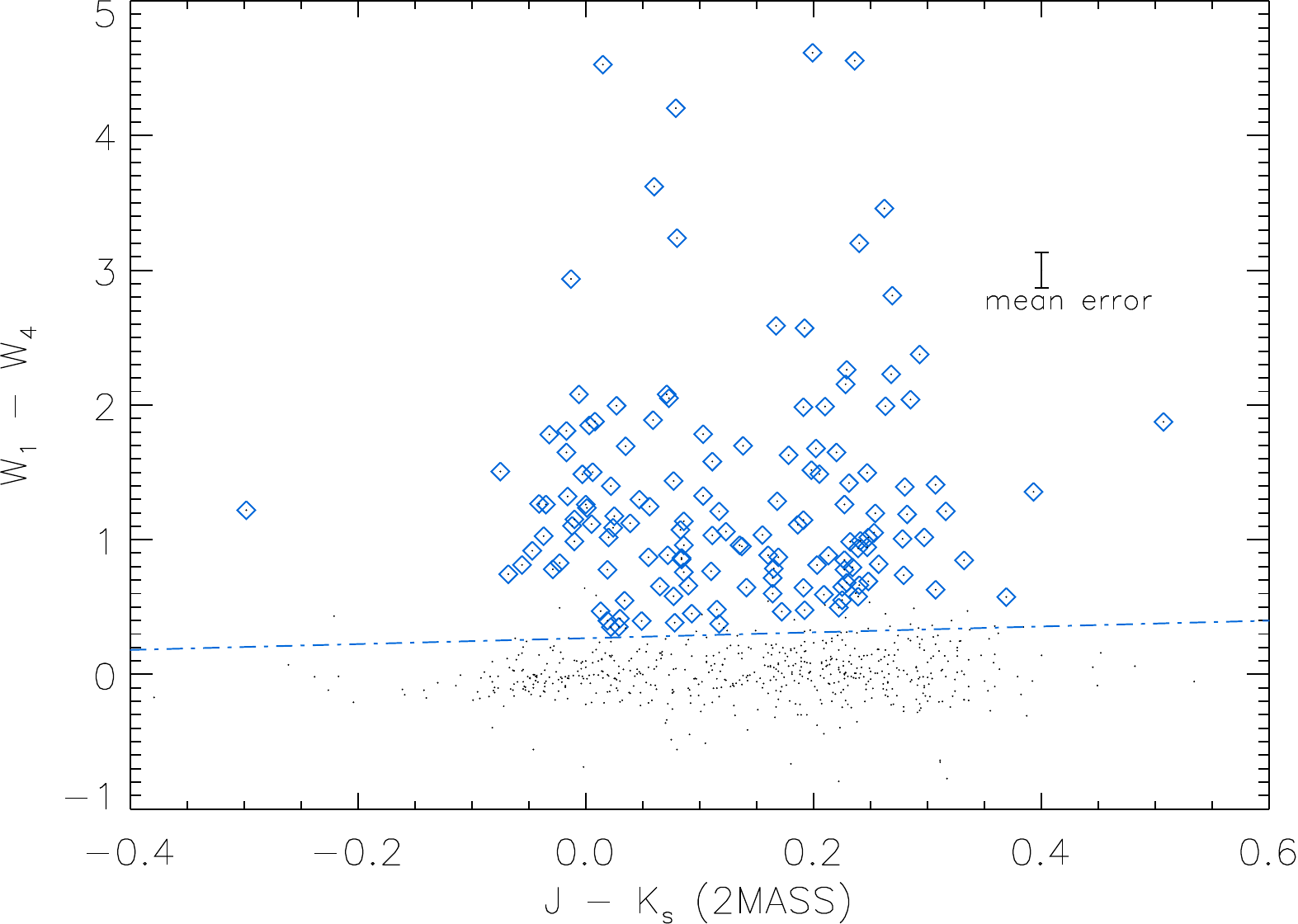}}
\caption{The colour-colour diagrams for the three WISE colours defined above. The blue line represents the corresponding excess detection threshold above which objects are considered to have a detectable excess. The photosphere grouping is clearly seen in each colour with fitted slopes and intercepts of (0.012,-0.18), (0.019,-0.018) and (0.22,-0.016) respectively. Blue diamonds indicate stars with detectable excesses.}
\label{figs}
\end{figure}

The criterion adopted for excess detection in the three colours was 8\% for \ws, 13\% for \whm~ and 30\% for \whl~ above the expected photosphere line, plus the error on the WISE photometry. These detection thresholds were chosen conservatively such that the detections are likely to be significant even if the photosphere fit is underestimated by 10\%. The colour-colour diagrams for the three WISE colours and the fits can be seen in Figure \ref{figs}.

Given excess detections in the three WISE colours we can remove from the sample those stars with suspect photometry. There were 19 objects with a detected excess in \ws~only, 3 in \whm~only, and 3 in \ws~and \whl~only. The 19 objects which show an excess only in the \ws~colour are likely caused by the saturation behaviour of the WISE photometry fits. When partial saturation occurs in band  W$_2$ (brighter than 6.5 mags), the WISE photometry fitting procedure produces a systematic over-estimation of the flux \citep{wise_sup}. All 19 of these objects are in the W$_2$ magnitude range where flux over-estimation is expected and hence we have treated these objects in our sample as having no detectable excess. In addition, we remove those objects with poor photometry fits ($\chi^2 > 4$) in the W$_1$ and W$_4$ bands, as the remainder of the analysis makes use of the \whl~ excess only; there were 95 such objects. HIP 77562 and 81891, which only show an excess in \whm~ both have poor photometry fits in band W$_4$,  and so are removed from the sample. HIP 77911 shows a poor photometry fit in the W$_1$ band ($\chi^2 = 4.6$), which is only marginally above our cut ($\chi^2 = 4$). HIP 77911 also shows a strong excess in the W$_2$-W$_4$ colour, where the photometry fits are reliable, and hence we have included HIP 77911 in our sample with a 22\,\mic excess.

The WISE images of the 173 objects with detected excesses were then visually inspected for the presence of close, unresolved companions and contamination from excess-producing nebulosity. This yielded 27 objects with excesses not likely to be caused by a disk. HIP 68413, 79098 and 76063, which have detectable excess in \ws~and \whl, but not \whm, are included in the sample based on the image inspection. The lack of \whm~excess associated with HIP 68413 is most likely due to a large error on \whm~and the detected \ws~excess associated with HIP 79098 was found to be caused by a clearly visible diffraction spike in the W$_2$ image. HIP 76063 has a W$_2$ magnitude of 5.6, and so the \ws~excess associated with this star is most likely due to saturation. Finally, HIP 80897 shows a nebulous excess in \whm~only, and so was removed from the sample. 
In total, we observe reliable excesses associated with 135 objects: 28 stars in US, 53 in UCL and 54 in LCC.

\section{Discussion}

\begin{figure*}
\subfloat[US Excess Fraction]{\label{exfrac_us}\includegraphics[width=0.4\textwidth]{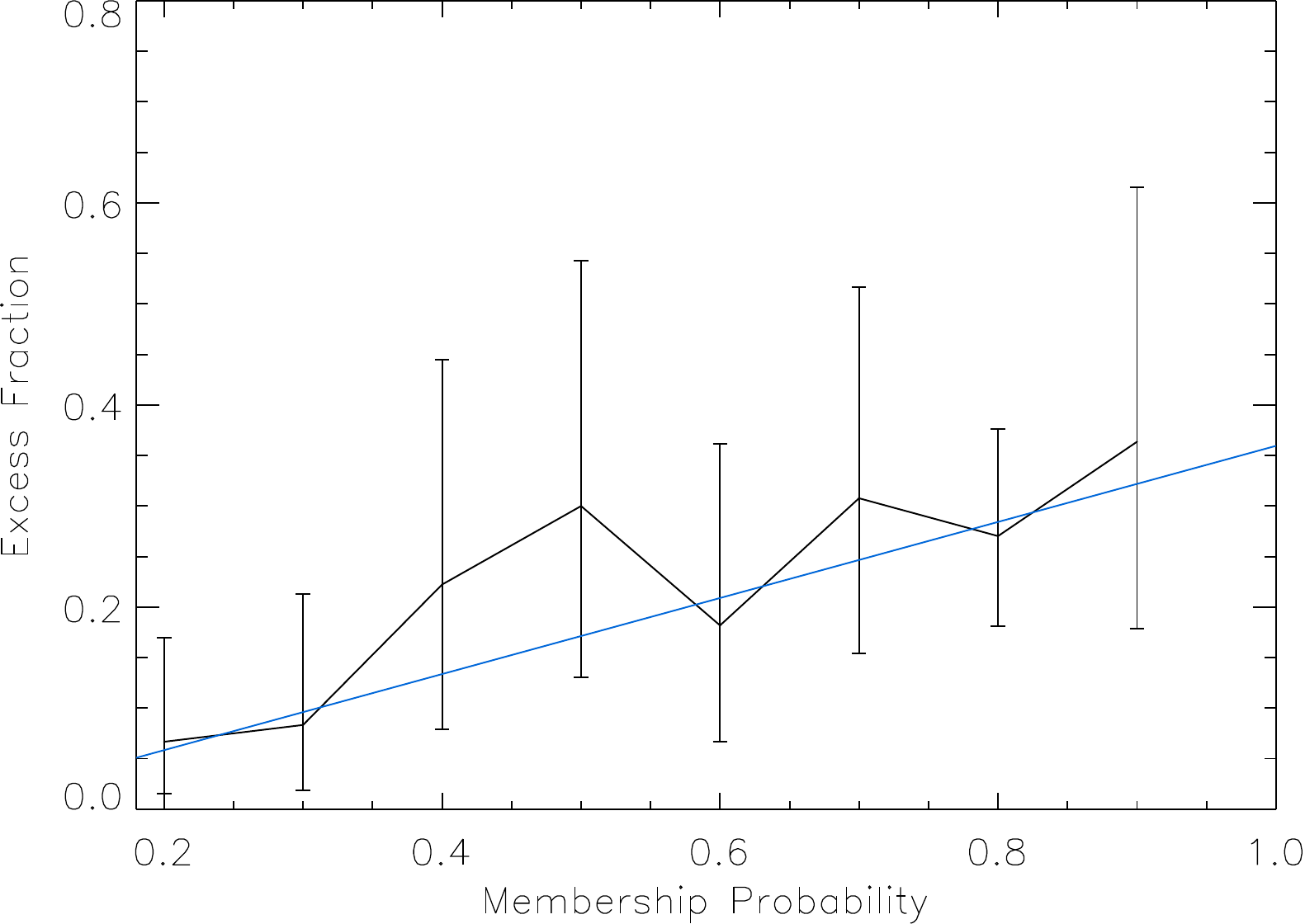}}
\subfloat[UCL Excess Fraction]{\label{exfrac_ucl}\includegraphics[width=0.4\textwidth]{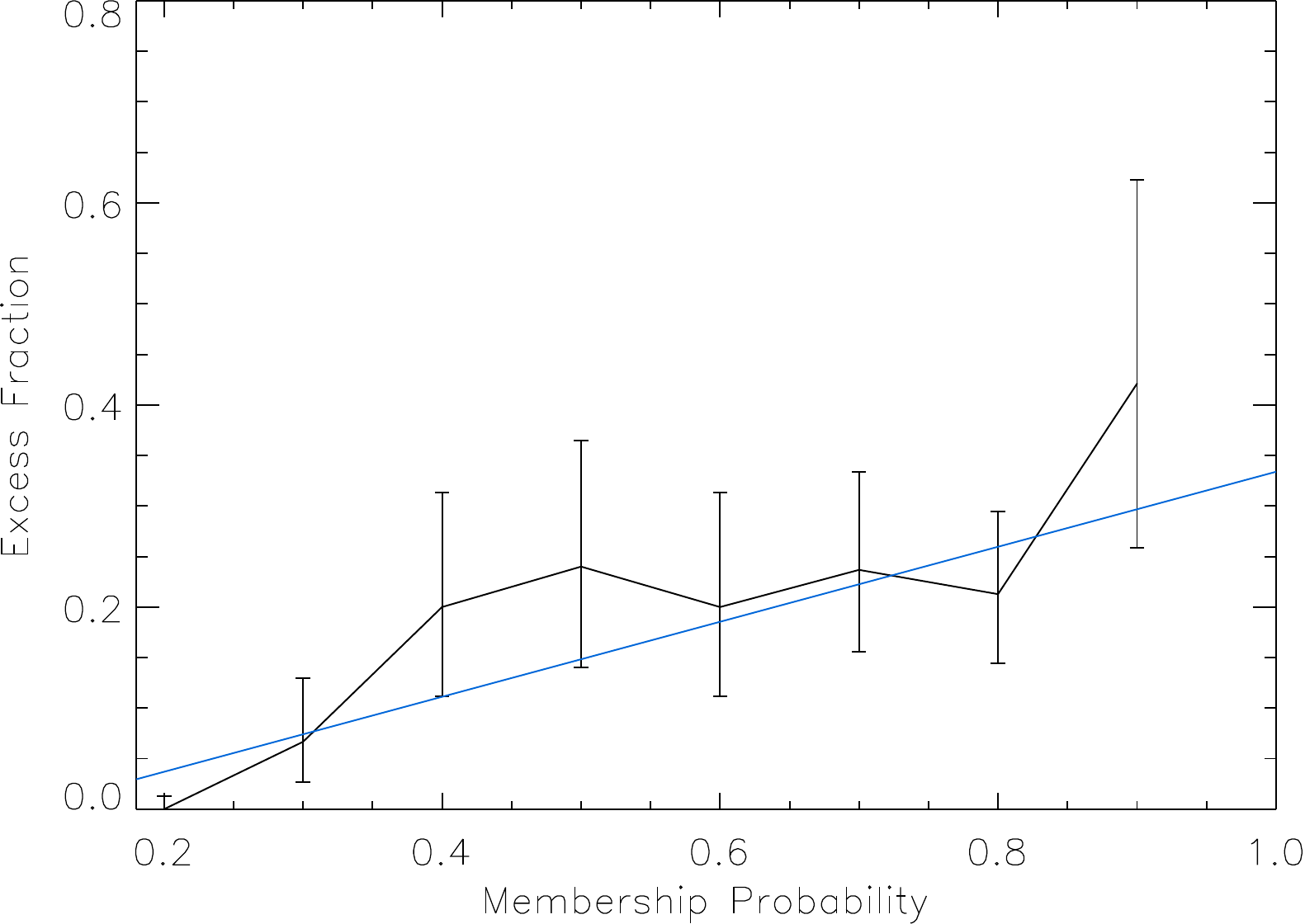}}\\
\subfloat[LCC Excess Fraction]{\label{exfrac_lcc}\includegraphics[width=0.4\textwidth]{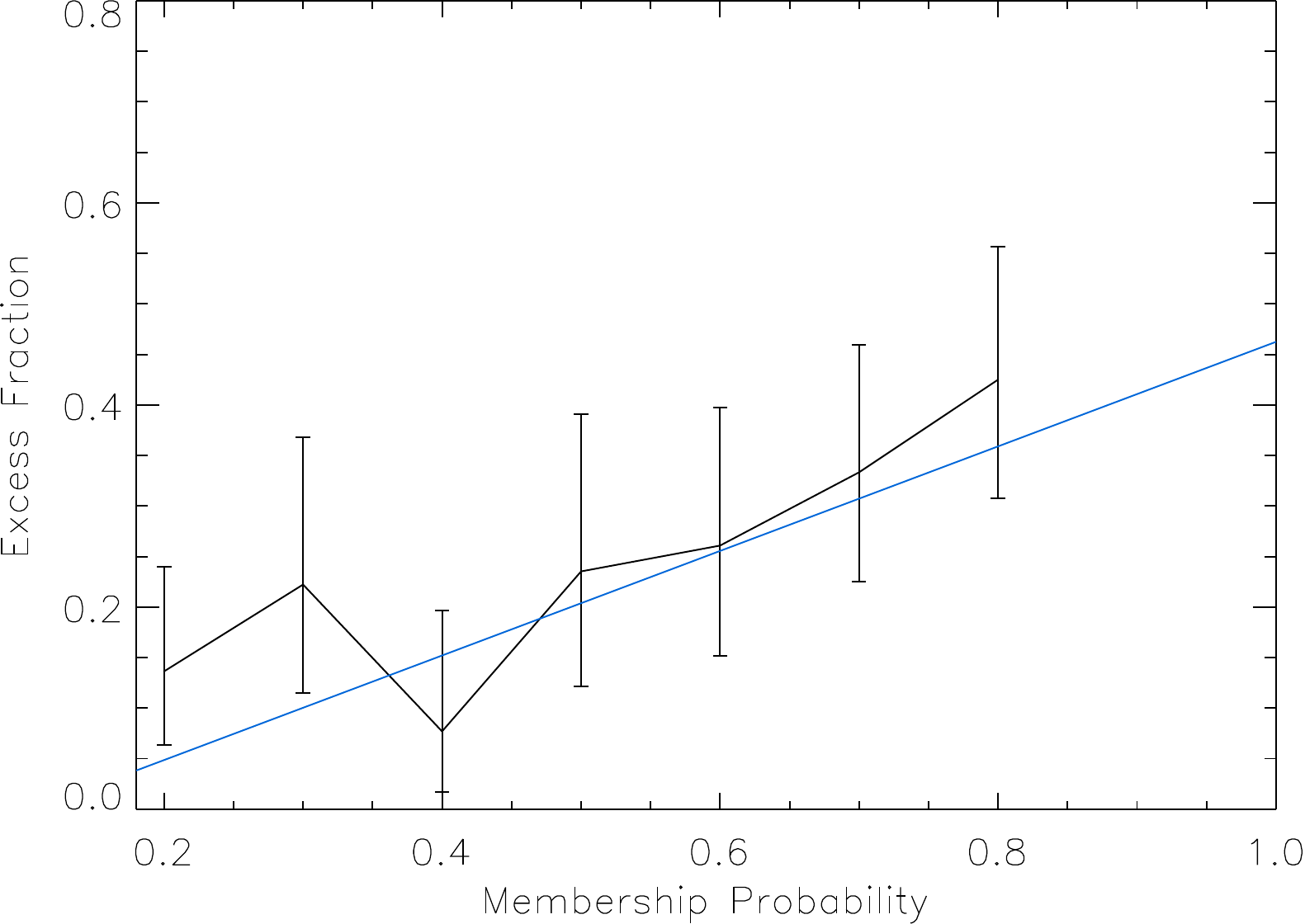}}
\subfloat[Sco-Cen Excess Fraction]{\label{exfrac_all}\includegraphics[width=0.4\textwidth]{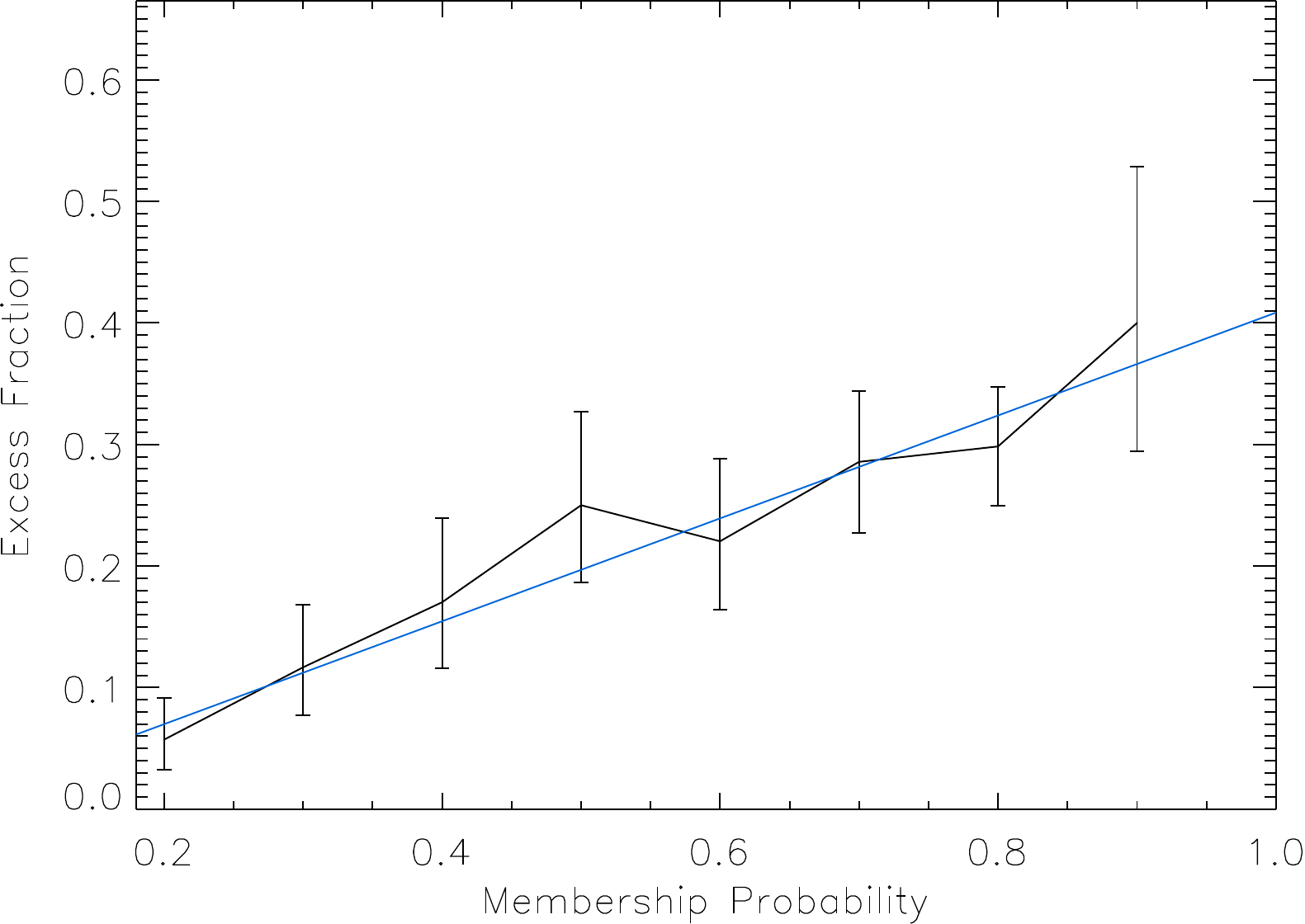}}
\caption{Excess fraction as a function of membership probability for the three subgroups and the entire association.} 
\label{exfigs}
\end{figure*}

The sources investigated in this study are BAF-type stars and hence the detected excesses are expected to be produced by dusty debris disks rather than primordial gaseous disks \citep{carpenter09}. A clear outcome of our analysis is that the excess fraction in the three subgroups is not uniform with respect to  membership probability (p). Figure \ref{exfigs} displays the excess fraction in 10\% membership probability bins with p $>$ 20\%. We have fitted linear trends to these data. Extrapolation to 100\% membership probability (i.e. certain members) along the linear fits results in excess fractions of 0.38$\pm$0.1, 0.33$\pm$0.08 and 0.46$\pm$0.13 for the three subgroups. 

{\color{red} The extrapolated excess fraction for US is larger than the observed 24\,\mic excess fraction in the US sample used by \citet{carpenter09}, which was found to be $\sim$0.3 for B7-A9 stars and $\sim$0.15 for F-type stars, and is in agreement with the value of $\sim$0.33 for F and G-type US stars seen by \citet{chen11}}. For associations at the age of UCL and LCC (16, 17\,Myr), \citet{chen05} reported a 24\,\mic excess fraction lower-bound of $\sim$0.35, and \citet{rieke05} and \citet{su06} observe that ~60\% of young stars ($<$30\,Myr) do not have an excess. These results are consistent with our study. We do not see an increase in the 22\,\mic excess fraction for the two older subgroups (UCL, LCC). Previous observations have provided some evidence for a peak in the excess fraction as stars age from $<$10\,Myr to the 10-30\,Myr age range \citep{currie08,gaspar09}. However, our analysis shows that in the Sco-Cen association, the excess fraction does not change significantly between the three subgroups. {\color{red}A recent study \citep{pecaut11} suggests that the true age of US may in fact be $\sim$11 Myr, which offers a potential explanation for the lack of an excess fraction increase between the subgroups.}

Given the lack of statistically significant differences in excess fraction between the three Sco-Cen subgroups, the association as a whole can be explored. Figure \ref{exfrac_all} displays the linear excess fraction trend for combined sample. The extrapolated excess fraction was 41$\pm$5\% for certain members, and consistent with 0\% for field stars, with a $\chi^2$ of 1.3 for the fit.

We have cross checked our sample with those of the Spitzer MIPS studies of \citet{chen11}, \citet{carpenter09}, \citet{su06}, and \citet{chen05} as a basic assessment of the relative success of the WISE photometry in identifying excesses. We see that of those stars in both data sets which passed our photometry tests, six have observed excesses above 30\% in the MIPS sample but are not assigned an excess in our study. Of these six stars, five have measured 22\,\mic flux as a multiple of the photosphere within one-sigma of the corresponding MIPS 24\,\mic values. The remaining star, HIP 79673, has a WISE flux ratio of 1.23$\pm$0.15, while the MIPS value is 1.53$\pm$0.05. \citet{chen11} indicate that the MIPS photometry for this star is contaminated, and the WISE image shows significant contamination in the form of well-resolved extended emission. This is most likely the cause of the discrepancy. Furthermore, there are two sources with large MIPS flux ratios (5-10 times the photosphere) which did not pass our photometry tests. HIP 61087 has very poor WISE photometry fits as well as clear contamination in the images (and in the MIPS data), though the measured flux ratio is similar to that of \citet{chen11}. The presence of an excess cannot be made clear from the data and so no further comments on this object can be made. Finally, HIP 80921 shows a relatively strong excess, however inspection of the images clearly show that the excess is caused by extended emission . Similarly, the Spitzer MIPS images show extended emission with no clear stellar source at 24\,\mic.

\begin{table*}
\begin{tabular}{c c c c c c c c c c c c}
\hline
HIP&P&$F_{W_4}$&$\sigma_{F_{w_4}}$&R($W_2$)& $\sigma_{R(W_2)}$ &R($W_3$)& $\sigma_{R(W_3)}$&R($W_4$)&$\sigma_{R(W_4)}$&Excess$^{a}$& Subgroup\\
\hline
 49360  &  6 &      12.02 &   0.76  &   0.99 &   0.03 &   0.98 &   0.03 &   1.03 &   0.07 &     NNN & LCC \\
 50612  &  7 &      14.02 &   0.80  &   0.99 &   0.03 &   1.05 &   0.03 &   1.46 &   0.09 &     NNY & LCC \\
 50667  &  7 &      11.54 &   0.71  &   1.01 &   0.03 &   1.23 &   0.04 &   2.24 &   0.15 &     NYY & LCC \\
\hline
\end{tabular}
\caption{{\bf N.B.} Complete table online. The  and WISE band fluxes as a multiple of the photosphere calculated from the three WISE colours and the 22\,\mic flux. $^{a}$ This column indicates the detection of an excess in the three colours. The fluxes are in mJy.}
\label{thetable}
\end{table*}

The correlation between the presence of a circumstellar disk and membership of the Sco-Cen association is clearly demonstrated in this analysis. It is thus important to consider stars with a detectable 22\,\mic excess which have lower membership probabilities (p$<$50\%). In the latest Sco-Cen membership study \citep{myfirstpaper} a number of \citet{zeeuw99} member stars were assigned low membership probabilities due to inconclusive proper motion data and the lack of a radial velocity measurement. Unresolved multiplicity has long been recognised as an important pitfall in kinematics-based association membership selection methods. An equal mass binary system at the distance of Sco-Cen can produce a proper-motion offset on the order of $\sim$2\,mas from the true centre-of-mass motion. This is approximately the size of the uncertainties in the proper motion \citep{leeuwen07}, indicating that binary association members can be overlooked. The presence of a circumstellar disk can then be used to indicate membership for stars spatially and photometrically consistent, but kinematically inconsistent, with the Sco-Cen subgroups. We thus increase the Bayes' factors (see Rizzuto et al. 2011) of stars with between 10\% and 50\% membership probability and a detected excess by a conservative factor of 4.8 (Table \ref{newmems}). This is the lower bound of the 99\% confidence interval of the ratio of excess fractions at 100\% and 0\% membership probability.  Note that including these stars in a sample of Sco-Cen members will necessarily introduce a bias toward disk presence.

\begin{table}
  \begin{tabular}{c c c c c c c c}
    \hline
    HIP&P&HIP&P&HIP&P&HIP&P\\
    \hline
    51169 & 57 & 51203 & 71 & 53524 & 81 & 55616 & 46\\
    60183 & 69 & 62482 & 59&  62488 & 72 & 63236 & 76\\
    63395 & 58 & 72099 & 81&   76223 & 81 & 76782 & 75\\
    77315 & 80 & 77523 & 70&    78198 & 76 & 78357 & 49  \\
    78826 & 79 &78943 & 82&    80019 & 64 &80458 & 82  \\
    81316 & 77 &82154 & 48 &  83232 & 36 &84881 & 43  \\
    86853 & 50 && \\
    \hline
  \end{tabular}
  \caption{The 25 excess detections around low probability members and the adjusted probabilities.}
  \label{newmems}
\end{table}

\begin{table}
\centering
  \begin{tabular}{c c c c c}
    \hline
    Figure&a&b&Corr&$\chi^2$\\
    \hline
    US  & 0.38$\pm$0.10 & -0.02$\pm$0.09& -0.65 &0.6\\
    UCL & 0.33$\pm$0.08 & -0.04$\pm$0.06& -0.68 &2.2\\
    LCC & 0.46$\pm$0.13 & -0.05$\pm$0.11& -0.77 &3.3\\
    All & 0.41$\pm$0.05 & -0.02$\pm$0.04& -0.65 &1.3\\
    B   & 0.18$\pm$0.10 &  0.09$\pm$0.10& -0.73 &0.5\\
    A   & 0.45$\pm$0.11 &  0.05$\pm$0.09& -0.61 &2.2\\
    F   & 0.47$\pm$0.09 & -0.08$\pm$0.04& -0.77 &2.8\\
\hline
  \end{tabular}
\caption{The excess fraction fits for the seven graphs. The fitting was done to the equation $y = (a-b)p_{mem} + b$, where a and b are the excess fractions at $p_{mem}=$ 1.0 and 0 respectively.}
\label{fitstab}
\end{table}

The subgroup LCC has an anomalous concentration of stars with detectable excess in the 5-10\% membership probability range. Three of these stars, HIP 50612, 52867 and 53992, are known members of the young open cluster IC2602, which is on the far side of LCC \citep{Robichon99}, and are thus expected to have low Sco-Cen membership probabilities. Confusion with young background sources for this subgroup, which is on the Galactic plane, further contributes to the anomalous high excess fraction.


\begin{figure}
\centering
\includegraphics[width=0.4\textwidth]{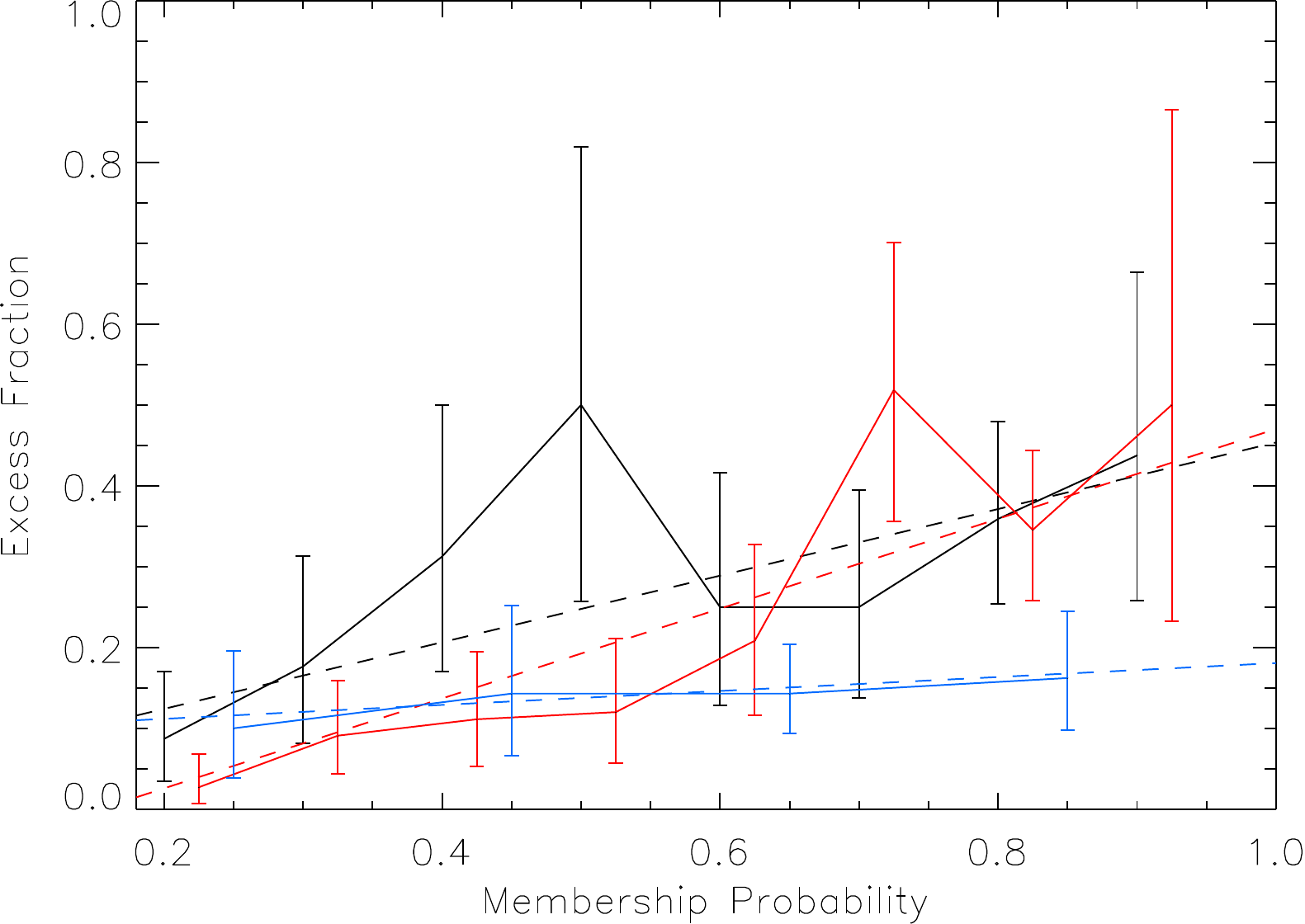}
\caption{Excess fraction against membership probability for A-type (Black), F-type (Red) and B-type (Blue) stars in our sample. The dashed lines represent the linear fits.}
\label{sptype_exfrac}
\end{figure}

We have examined the excess fraction properties of our sample in three colour ranges: ($-0.3<$B-V$<0$), ($0<$B-V$<0.3$) and ($0.3<$B-V$<0.6$). These groupings correspond approximately to B, A and F-type stars according to the colour tables of \citet{aaq2000}. Figure \ref{sptype_exfrac} displays plots of the excess fraction against membership probability for the three spectral type ranges. We find the extrapolated excess fractions for the A and F-type stars to be 45$\pm$11\% and 47$\pm$9\%, while the B-type stars in our sample show no evidence of a trend in excess fraction with membership probability. The study of \citet{carpenter09} also found that A and F-type stars have similar excess fractions at the age of Sco-Cen, with a small increase in excess fraction for the earlier spectral types. However, the earliest star included in the \citet{carpenter09} sample is B7, and so no direct comparison can be made to the bluest end of our sample. 


\section{Summary and Conclusions}
We have analysed the available preliminary WISE photometry for the Sco-Cen stars of the \citet{myfirstpaper} membership list and detected 135 22\,\mic excesses above the expected photosphere emission. We have used Sco-Cen membership probabilities to extrapolate an excess fraction for certain members, {\color{red} and observe that there is no significant difference in excess fraction between the three association subgroups. This agrees with previous studies \citep{carpenter09} and may be explained by the revised US age of $\sim$11\,Myr \citep{pecaut11}.} Importantly we find that the excess fraction is significantly lower for the B-type stars in our sample compared to A and F-type association members, which is contrary to the trend seen by \citet{carpenter09}. It is possible that the lack of a clear trend for the B-type sample could indicate that most of them are in fact young Sco-Cen members despite their kinematics. Another possible explanation relates to multiplicity. B-type stars have a significantly higher multiplicity fraction compared to later type stars \citep{kouwenhoven05}. The presence of a companion can potentially truncate the inner regions of the circumstellar disk through resonances \citep{lubow94}, producing a smaller disk fraction. This has been observed in a recent study of binaries in Taurus-Auriga, particularly with close ($<$40\,AU) companions \citep{kraus11arxiv}. Among the highest probability members in our sample ($>$90\%) there are six close ($<$100\,AU) multiple systems without disk detections and one close multiple system with a detected excess. A closer, more comprehensive, comparison between disk presence and multiplicity information for the Sco-Cen B-type stars may shed light on this issue, but is beyond the scope of this study.

\bibliography{references}
\bibliographystyle{mn2e}

\onecolumn
\begin{longtable}{c c c c c c c c c c c c}
\caption{The WISE band 4 flux and $W_2$,$W_3$ and $W_4$ flux as a multiple of the photosphere with errors for our sample. $^{a}$ This column indicates the detection of an excess in the three colours. The lower section of the table displays the objects removed from our sample due to poor photometry or unreliable images.}\\
\hline
HIP&P&$F_{W_4}$ (mJy)&$\sigma_{F_{w_4}}$ (mJy)&R($W_2$)& $\sigma_{R(W_2)}$ &R($W_3$)& $\sigma_{R(W_3)}$&R($W_4$)&$\sigma_{R(W_4)}$&Excess$^{a}$& Subgroup\\
\hline
\\
\endhead
 49360  &  6 &      12.02 &   0.76  &   0.99 &   0.03 &   0.98 &   0.03 &   1.03 &   0.07 &     NNN & LCC \\
 50612  &  7 &      14.02 &   0.80  &   0.99 &   0.03 &   1.05 &   0.03 &   1.46 &   0.09 &     NNY & LCC \\
 50667  &  7 &      11.54 &   0.71  &   1.01 &   0.03 &   1.23 &   0.04 &   2.24 &   0.15 &     NYY & LCC \\
 51169  & 22 &       8.61 &   0.72  &   1.00 &   0.03 &   1.25 &   0.04 &   2.17 &   0.19 &     NYY & LCC \\
 51203  & 33 &       9.97 &   0.74  &   0.99 &   0.03 &   1.06 &   0.03 &   1.55 &   0.12 &     NNY & LCC \\
 51507  &  9 &      11.47 &   0.92  &   1.02 &   0.02 &   0.98 &   0.02 &   0.90 &   0.07 &     NNN & LCC \\
 51508  &  6 &      12.72 &   0.82  &   0.97 &   0.03 &   0.97 &   0.03 &   1.04 &   0.07 &     NNN & LCC \\
 52059  &  6 &      13.41 &   0.85  &   0.98 &   0.03 &   0.99 &   0.03 &   0.94 &   0.07 &     NNN & LCC \\
\end{longtable}

\end{document}